\documentclass[a4paper,english]{aa}
\usepackage{times}
\usepackage[T1]{fontenc}
\usepackage[latin1]{inputenc}
\usepackage{amsmath}
\usepackage{graphicx}
\usepackage{amssymb}
\usepackage[authoryear]{natbib}

\makeatletter

\providecommand{\tabularnewline}{\\}

\def\jnl@style{\it}
\def\aaref@jnl#1{{\jnl@style#1}}

\def\aaref@jnl#1{{\jnl@style#1}}

\def\aj{\aaref@jnl{AJ}}                   
\def\araa{\aaref@jnl{ARA\&A}}             
\def\apj{\aaref@jnl{ApJ}}                 
\def\apjl{\aaref@jnl{ApJ}}                
\def\apjs{\aaref@jnl{ApJS}}               
\def\ao{\aaref@jnl{Appl.~Opt.}}           
\def\apss{\aaref@jnl{Ap\&SS}}             
\def\aap{\aaref@jnl{A\&A}}                
\def\aapr{\aaref@jnl{A\&A~Rev.}}          
\def\aaps{\aaref@jnl{A\&AS}}              
\def\azh{\aaref@jnl{AZh}}                 
\def\baas{\aaref@jnl{BAAS}}               
\def\jrasc{\aaref@jnl{JRASC}}             
\def\memras{\aaref@jnl{MmRAS}}            
\def\mnras{\aaref@jnl{MNRAS}}             
\def\pra{\aaref@jnl{Phys.~Rev.~A}}        
\def\prb{\aaref@jnl{Phys.~Rev.~B}}        
\def\prc{\aaref@jnl{Phys.~Rev.~C}}        
\def\prd{\aaref@jnl{Phys.~Rev.~D}}        
\def\pre{\aaref@jnl{Phys.~Rev.~E}}        
\def\prl{\aaref@jnl{Phys.~Rev.~Lett.}}    
\def\pasp{\aaref@jnl{PASP}}               
\def\pasj{\aaref@jnl{PASJ}}               
\def\qjras{\aaref@jnl{QJRAS}}             
\def\skytel{\aaref@jnl{S\&T}}             
\def\solphys{\aaref@jnl{Sol.~Phys.}}      
\def\sovast{\aaref@jnl{Soviet~Ast.}}      
\def\ssr{\aaref@jnl{Space~Sci.~Rev.}}     
\def\zap{\aaref@jnl{ZAp}}                 
\def\nat{\aaref@jnl{Nature}}              
\def\iaucirc{\aaref@jnl{IAU~Circ.}}       
\def\aplett{\aaref@jnl{Astrophys.~Lett.}} 
\def\apspr{\aaref@jnl{Astrophys.~Space~Phys.~Res.}}
\def\bain{\aaref@jnl{Bull.~Astron.~Inst.~Netherlands}} 
\def\fcp{\aaref@jnl{Fund.~Cosmic~Phys.}}  
\def\gca{\aaref@jnl{Geochim.~Cosmochim.~Acta}}   
\def\grl{\aaref@jnl{Geophys.~Res.~Lett.}} 
\def\jcp{\aaref@jnl{J.~Chem.~Phys.}}      
\def\jgr{\aaref@jnl{J.~Geophys.~Res.}}    
\def\jqsrt{\aaref@jnl{J.~Quant.~Spec.~Radiat.~Transf.}}
\def\memsai{\aaref@jnl{Mem.~Soc.~Astron.~Italiana}}
\def\nphysa{\aaref@jnl{Nucl.~Phys.~A}}   
\def\physrep{\aaref@jnl{Phys.~Rep.}}   
\def\physscr{\aaref@jnl{Phys.~Scr}}   
\def\planss{\aaref@jnl{Planet.~Space~Sci.}}   
\def\procspie{\aaref@jnl{Proc.~SPIE}}   

\usepackage{babel}
\makeatother
\begin{document}

\title{The superburst recurrence time in luminous persistent LMXBs}

\author{L.~Keek \inst{1,2,4}, J.~J.~M.~in~'t~Zand\inst{1,2}\and A.~Cumming\inst{3}}

\institute{SRON Netherlands Institute for Space Research, Sorbonnelaan 2, 3584
CA Utrecht, The Netherlands \and  Astronomical Institute, Utrecht
University, Princetonplein 5, 3584 CC Utrecht, The Netherlands\and Physics
Department, McGill University, 3600 rue University, Montreal, QC,
H3A 2T8, Canada\and \email{l.keek@sron.nl}}

\date{Received ; Accepted}

\abstract{Theory and observations favor stable helium burning as the most important
means to produce fuel for superbursts on neutron star surfaces. However,
all known superbursters exhibit unstable burning as well. This ambiguity
prompted us to search for superbursts in data from the BeppoSAX Wide
Field Cameras of ten luminous LMXBs, most of which do not exhibit
normal type-I X-ray bursts. We found no superbursts and determine
a lower limit on the recurrence time which varies between 30 and 76
days (90\% confidence). All recurrence time limits except one are
longer than the observed recurrence time for GX~17+2. This difference
can be understood if the mass accretion rate in GX~17+2 is several
tens of percent higher than in the other sources; alternatively, the
accreted material in GX\textasciitilde{}17+2 might be hydrogen deficient,
leading to larger carbon yields than in the other sources. We compare
our results to the latest models of superbursts. As our search method
is indiscriminate of the burst ignition scenario, the recurrence time
limits may also be applied to other bursts of similar duration and
brightness.
}

\maketitle

\section{Introduction}

Neutron stars in low-mass X-ray binaries (LMXBs) accrete hydrogen-
and helium-rich matter from a sub-solar mass companion star. The gravitational
energy that is released when the matter travels to the surface of
the neutron star accounts for most of the observable X-ray radiation.
The accreted matter forms a layer on the neutron star. When this layer
is sufficiently compressed and heated, thermonuclear burning is ignited.
The burning can be stable or unstable, depending on e.g. the mass
accretion rate. Unstable nuclear burning results in a type I X-ray
burst (\citealt{1976Grindlay,1976Belian,1976Woosley,1977Maraschi}).
A burst is observed as a fast (1--10 seconds) rise in the flux and
a slow (10--1000 seconds), exponential-like decay. Many such bursts
have been observed from about 80 sources (see e.g. \citealt{2004ZandWFC}).
For reviews see \citet{1993Lewin} and \citet{Strohmayer:2003vf}.

Longer X-ray bursts have been observed that have a similar fast rise,
but a much longer decay of several hours. These so-called superbursts
release about a thousand times more energy than the normal type I
X-ray bursts. So far, ten superbursters have been identified, with
two exhibiting more than one superburst. For an observational overview,
see \citet{Kuulkers:2003wt} (and see also \citealt{2004intZand},
\citealt{2005ATel..482....1R} and \citealt{2005ATel..483....1K}
for reports of new superbursters).

Superbursts are ascribed to unstable carbon burning (initiated by
$^{12}\mathrm{C}+{}^{12}\mathrm{C}\rightarrow\mathrm{^{20}Ne}+\alpha$)
in an `ocean' of heavy elements which lies between the freshly accreted
layer and the neutron star crust (\citealt{2001CummingBildsten},
\citealt{2002Strohmayer}). The carbon and heavy elements are thought
to be the products of rp-process hydrogen and helium burning in the
accreted layer (see e.g. \citealt{1999Schatz,2003Schatz}). This model
has been fairly successful at reproducing the energetics and recurrence
times of superbursts (\citealt{2001CummingBildsten,2002Strohmayer,2004Brown,2005CooperNarayan,2005Cumming1}),
as well as the observed lightcurves and the quenching of normal type
I X-ray bursts for about a month following a superburst (\citealt{2004CummingMacBeth,2005Cumming1}).

An open issue is how to produce enough carbon to power a superburst.
Carbon fractions of more than 10\% are required to reproduce the observed
lightcurves of superbursts and to achieve unstable ignition at the
accretion rates of approximately $0.1\dot{M}_{\mathrm{Edd}}$ inferred
for most of the superbursters ($\dot{M}_{\mathrm{Edd}}$ is the Eddington
accretion rate). Calculations of rp-process burning show that stable
rather than unstable H/He burning is required for such large carbon
fractions (\citealt{2003Schatz}). The reason is that carbon production
is only possible by helium burning after the hydrogen has been consumed,
since hydrogen readily captures on carbon, processing it to heavy
elements. Helium is rapidly consumed at the high temperatures reached
during unstable burning. However, in stable burning, the helium burns
at a reduced rate so that when the hydrogen has been completely consumed
by rp-process burning, there is still some helium left that then burns
into carbon. In this way, there is an anti-correlation between the
amount of carbon that is produced and the length of the rp-process,
or equivalently the average mass of the heavy elements (\citealt{2003Schatz}).

Observations support the presence of stable burning in superburst
sources. In~'t~Zand~et~al. \citeyearpar{2003intZand} showed that
superburst sources preferentially have large values of alpha, the
ratio of the persistent fluence between type I X-ray bursts to the
fluence in the type I X-ray burst. Large values of alpha indicate
that not all the nuclear fuel is consumed in bursts, and that stable
burning must occur between them. \citet{2003Cornelisse} studied BeppoSAX
observations of nine type I X-ray bursters and found that the burst
rate drops for luminosities exceeding approximately 10\% of the Eddington
limit, implying that stable burning is occurring. This critical luminosity
is close to the lowest luminosity of superburst sources, and may explain
why superbursts have not been seen at accretion rates less than $\sim0.1\dot{M}_{\mathrm{Edd}}$.

In contrast, the role of unstable burning and the heavy elements produced
by the rp-process remains unclear. Observationally, it is true that
all superburst sources show type I X-ray bursts. \citet{2001CummingBildsten}
pointed out that for a fixed heat flux emerging from the neutron star
crust, the heavy nuclei reduce the thermal conductivity of the accumulating
ocean, leading to higher temperatures and earlier carbon ignition
than previously found in models of pure carbon layers (see e.g. \citealt{1976Woosley}).
However, \citet{2004Brown} showed by considering the thermal state
of the entire neutron star that in fact the thermal state of the core
and crust of the neutron star is most important for setting the ignition
column%
\footnote{This is a very exciting result because it opens up the possibility
of using superburst observations as a probe of the neutron star interior
(\citealt{2004Brown,2005CooperNarayan,2005Cumming1}).%
}. Because the flux emerging from the crust changes with the heavy
element composition, the overall effect of the heavy elements is much
smaller than calculated by \citet{2001CummingBildsten}. Nonetheless,
the heavy elements do have some effect on the ignition depth, and
so it is possible that unstable burning could contribute to the presence
of superbursts.

One way to address this issue is to study the several persistently
bright LMXBs which either do not or very rarely show type I X-ray
bursts. The rapidly accreting source GX~17+2 has already been important
for testing carbon ignition models. \citet{2001CummingBildsten} showed
that superbursts should occur at rapid accretion rates near Eddington,
but with reduced energies and recurrence times. In~'t~Zand~et~al.~\citeyearpar{2004intZand}
identified four superbursts from GX~17+2 which well-matched the predicted
recurrence times, although their durations were longer than expected.
In this paper, we report on a systematic search for superbursts in
BeppoSAX Wide Field Camera data of ten LMXBs with sufficiently high
accretion levels. This includes the so-called {}``GX'' sources:
very bright LMXBs with luminosities close to the Eddington limit.
We include six sources that never exhibited an X-ray burst as well
as four X-ray bursters for comparison purposes, including the known
superburster GX~17+2. We compare the behavior of the four most luminous
sources Sco~X-1, GX~340+0, GX~5-1 and Cyg~X-2 to the equally luminous
superburster GX~17+2. For all these sources we find a constraining
lower limit on the superburst recurrence time.

\section{Observations\label{sec:Observations}}

We search for superbursts in archival data from the Wide Field Cameras
(WFC; \citealt{1997Jager}) on BeppoSAX (\citealt{1997Boella}) of
ten X-ray sources. The two WFCs were identical coded mask telescopes
pointing in opposite directions. They had $40^{\circ}\times40^{\circ}$
fields of view and a $5^{\prime}$ angular resolution in the 2--25~keV
bandpass. They were active between 1996 and 2002. The WFCs carried
out a program of semi-yearly campaigns on the Galactic center region.
For a review see \citet{2004ZandWFC}. We consider nine sources near
the Galactic center and Cyg~X-2 (see Table \ref{cap:Results-for-9}).
The selected sources accrete mass at a comparable rate to the known
superbursters. Some of the sources exhibit normal bursts while others
do not.

For each source there is 55 to 153 days of observation time accumulated
over a period of six years (see Table \ref{cap:Results-for-9}). Note
that this is not the net exposure time, but includes data gaps due
to Earth occultations (typically 36 minutes long and recurring every
96 minutes) and passages through the South Atlantic Anomaly (13 to
26 minutes long). During these data gaps there is no observation of
the source. However, the duration of a data gap is short with respect
to the expected duration of a superburst. These data gaps are, therefore,
not very detrimental to the detection of superbursts. Except for these
short datagaps, an uninterrupted period of exposure time is typically
1.5 days long.

To determine the persistent flux in the 0.1--200~keV range of GX~9+9
we use data from the BeppoSAX Narrow-Field Instruments (NFI). Of these
instruments we use data from the Low Energy Concentrator Spectrometer
(LECS; \citealt{1997Parmar}; 0.1--10~keV), the Medium Energy Concentrator
Spectrometers two and three (MECS; \citealt{1997BoellaMECS}; 1.3--10~keV)
and the Phoswich Detection System (PDS; \citealt{1997Frontera}; 13--200~keV).
For the other sources we obtain a measure of the flux from the literature
(see Section \ref{sec:Mass-accretion-rate}). \begin{table*}

\caption{\label{cap:Results-for-9}Superburst recurrence time of ten sources.
{\small For each source we give the time in days between the first
and the last observation $t_{\mathrm{span}}$, the total observation
time $t_{\mathrm{obs}}$ including data gaps, the flaring fraction,
the number of observed superbursts (SBs), whether a source exhibits
normal bursts and the (lower limit of the) superburst recurrence time
$t_{\mathrm{recur}}$.}}

\begin{center}\begin{tabular}{llllccl}
\hline 
Source&
$t_{\mathrm{span}}${[}d{]}&
$t_{\mathrm{obs}}${[}d{]}&
flaring fraction&
\# SBs&
burster&
$t_{\mathrm{recur}}${[}d{]}\tabularnewline
\hline
\object{Sco X-1}&
1870.03&
55.44&
16\%&
0&
no&
>30\tabularnewline
\object{GX 340+0}&
2062.82&
153.20&
10\%&
0&
no&
>74\tabularnewline
\object{GX 349+2}&
2062.82&
129.71&
12\%&
0&
no&
>66\tabularnewline
\object{GX 9+9}&
2062.57&
139.27&
8\%&
0&
no&
>76\tabularnewline
\object{GX 354-0}&
2062.79&
100.97&
10\%&
0&
yes&
>52\tabularnewline
\object{GX 5-1}&
2062.79&
109.95&
11\%&
0&
no&
>55\tabularnewline
\object{GX 9+1}&
2062.79&
103.45&
9\%&
0&
no&
>53\tabularnewline
\object{GX 13+1}&
2062.79&
104.30&
9\%&
0&
yes&
>55\tabularnewline
\object{GX 17+2}&
2062.80&
127.59&
16\%&
4&
yes&
$30\pm15^{\mathrm{a}}$\tabularnewline
\object{Cyg X-2}&
1999.10&
109.67&
11\%&
0&
yes&
>60\tabularnewline
\hline
\end{tabular}\end{center}

{\small $\mathrm{^{a}}$ From \citet{2004intZand}.} \end{table*}

\section{Search for superbursts\label{sec:Search-for-superbursts}}

For each source we have 55 to 153 days of observation time while a
superburst is typically observable for several hours. From the superbursts
thus far detected we know that they are rare, so we expect to find
at most a few new superbursts. The search for these consists of an
automatic peak-search algorithm and a visual inspection of the data.

We use a lightcurve in the full bandpass at three different time resolutions.
The highest time resolution is $16\,\mathrm{s}$. This resolution
allows to discriminate between fast and slow rising peaks. A higher
time resolution is not useful because the uncertainty in each lightcurve
point becomes too large. The next time resolution is 0.02 days (about
0.5 hr). We find this resolution good to resolve superbursts at optimum
sensitivity since short features such as normal bursts are averaged
out while superbursts are not. The lowest time resolution is $0.1\,\mathrm{d}$.
We employ it to measure the non-flaring persistent flux. At this resolution
superbursts are largely averaged out while typical variations in the
persistent flux are well sampled.

The search algorithm proceeds as follows. At $0.02\,\mathrm{d}$ time
resolution we check each data point whether it exceeds the persistent
flux significantly. Each data point is assigned a measure of the persistent
flux through a linear interpolation of the preceding and following
$0.1\,\mathrm{d}$-resolution data point. We then check whether the
$0.02\,\mathrm{d}$-resolution data point exceeds the persistent flux
at $4.4\sigma$ significance. We choose $4.4\sigma$ since at this
level of significance we detect all superbursts of GX~17+2 while
limiting the number of false peaks. There are typically 6000 points
in a $0.02\,\mathrm{d}$-resolution lightcurve. We expect the number
of false peaks exceeding $4.4\sigma$ due to Gaussian noise to be
0.06
.

Using the lightcurve at $16\,\mathrm{s}$ time resolution we check
these peaks visually for superburst characteristics: a fast rise and
a slow, exponential-like decay.

The lightcurve may contain features that have the characteristics
of a superburst but are flares (\citealt{2004intZand}). Flares are
thought to result from quick changes in the mass accretion rate. They
are generally not expected to be isolated events, but to come in multiples
during flaring episodes. Therefore we exclude obvious flaring periods.
We identify flaring episodes visually from the light curve. The fraction
of the lightcurve that shows flaring is presented in Table \ref{cap:Results-for-9}.
For the identification of a flaring episode we use a similar argument
as \citet{2004intZand}: an interval between two consecutive data
points at $0.1\,\mathrm{d}$-resolution is identified as a flaring
episode when the root mean square of the count rate at $0.02\,\mathrm{d}$
time resolution in that interval exceeds the average root mean square
of all intervals by a factor of two. The flaring fractions are on
average 11\%.

Using this method we reproduce the four superbursts found previously
for GX~17+2 as well as the two peaks that were determined to be flares
(\citealt{2004intZand}). For the other sources we find no superbursts
and we determine a lower limit on the superburst recurrence time $t_{\mathrm{recur}}$.

\section{Recurrence time\label{sec:Recurrence-time}}

Thus far only two out of ten superbursters exhibited more than one
superburst. These are therefore the only sources for which a recurrence
time has been directly obtained. From GX~17+2 four superbursts have
been observed with an average recurrence time of $30\pm15$ days and
from 4U~1636-536 three superburst were observed with recurrence times
of 1.75 and 2.9 years (\citealt{2004Kuulkers}).

To determine the lower limit of the recurrence time for the other
nine sources here investigated, we perform Monte Carlo simulations
of series of superbursts and cross check those with the WFC observation
schedule. We generate the series of onset times, for a given $t_{\mathrm{recur}}$,
by first randomly picking the time of the initial superburst between
0 and $t_{\mathrm{recur}}$. Subsequent superburst times are sampled
from a Gaussian distribution with a standard deviation $\sigma$,
centered around a time $t_{\mathrm{recur}}$ after the previous superburst.
We use the Gaussian distribution to model the variability in the recurrence
time resulting from the variability of the mass accretion rate as
measured through the flux. For the nine sources we find from the normalized
root mean squared of the flux that $\sigma$ is between 0.06 and $0.3\, t_{\mathrm{recur}}$.

The superburst exponential decay times are randomly sampled between
1 and 6 hours (see review by \citealt{Kuulkers:2003wt}). As we will
show in more detail below, we can take the decay time independently
from the wait time since the previous superburst, because we consider
decay times longer than the duration of the data gaps. We assume the
peak flux to be equivalent to half the Eddington limit of each source.
This implies that any superburst is detectable with the WFC for as
long as the e-folding decay time.

A given superburst is considered to be detected if at any time of
its e-folding decay duration the instrument was observing the source
according to the WFC observation schedule, taking into account all
data gaps, including those due to Earth occultations and passages
through the South Atlantic Anomaly, as well as flaring episodes.

For each source and $t_{\mathrm{recur}}$ we perform 100,000 Monte
Carlo simulations. With this number of simulations one expects an
uncertainty of $0.3$\% in the obtained value. We determine the percentage
of simulations in which at least one superburst is detected. We define
the lower limit to the recurrence time to be that value of $t_{\mathrm{recur}}$
for which there is a 90\% probability of detecting a superburst. The
results are provided in Table \ref{cap:Results-for-9}. We find lower
limits for $t_{\mathrm{recur}}$ between 30 and 76 days.

\begin{figure}
\begin{center}\includegraphics[%
  bb=90bp 250bp 522bp 520bp,
  clip,
  width=0.50\textwidth]{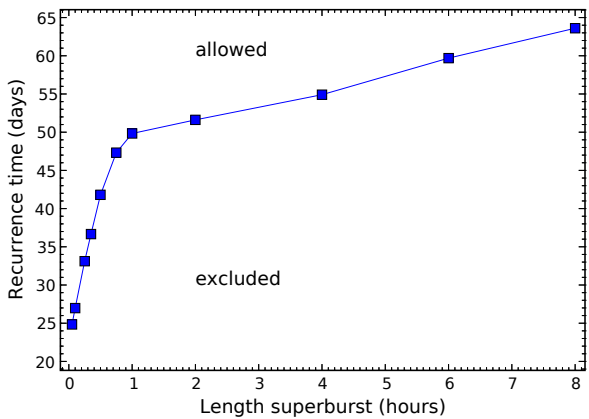}\end{center}

\caption{\label{cap:Lower-limit-to}Lower limit to the recurrence time at
90\% confidence as a function of assumed superburst duration in Monte
Carlo simulations. The lower limit drops steeply when the burst duration
is of the order of the length of the data gaps or less. Note that
our search is insensitive to bursts shorter than 0.5 hour due to our
choice of lightcurve time resolution (see Section \ref{sec:Search-for-superbursts}).}
\end{figure}

In Figure \ref{cap:Lower-limit-to} we illustrate the dependence of
the superburst detection on the duration from the Monte Carlo simulations.
For this we extend the duration range to below 1 hour. This clearly
shows that for superburst durations above 1 hour, the presence of
short datagaps is not detrimental to the detection of superbursts.%

Our aim is to look for superbursts, but in fact our search is not
limited to carbon flashes. The search method we use is indiscriminate
of the ignition scenario. Other X-ray bursts of sufficient duration
and brightness will also be found using our method. Long duration
bursts have been observed from several sources (see \citealt{2005Cumming1}
and \citealt{2002Kuulkers}). However, the exponential decay time
of these bursts is $\leq600\,\mathrm{s}$, which is shorter than the
lightcurve time resolution of 0.02d that we use (see Section \ref{sec:Search-for-superbursts}).
Therefore we would not have been able to detect these bursts. Note
that if a lightcurve at a shorter time resolution is used, such that
these long burst are detectable, the limits that can be placed on
the recurrence time are far less constraining than for superbursts,
since the burst duration is in this case shorter than the data gaps
in our observations (see Figure \ref{cap:Lower-limit-to}). Nevertheless,
the constraints we find on the recurrence time apply not only to superbursts,
but to any type of thermonuclear flashes which are long and bright
enough to be detected.

\section{Mass accretion rate\label{sec:Mass-accretion-rate}}

The unstably burning carbon layer to which superbursts are ascribed
is formed by the burning of accreted hydrogen and helium. The rate
at which the hydrogen and helium are accreted from the companion star
plays an important role in whether they burn stably or unstably. \begin{table*}

\caption{\label{cap:Results-for-non-bursting}Flux and luminosity estimates
of the 10 investigated sources. {\small We give the average persistent
flux $F$ in the energy range indicated below, the relative variability
of the flux as determined from ASM/RXTE measurements, distance $d$,
luminosity $L$ in units of the Eddington luminosity $L_{\mathrm{Edd}}\equiv\mathrm{2\cdot10^{38}erg\, s^{-1}}$
and the ignition column depth $y$. We indicate the uncertainty when
given, otherwise we assume a 2\% uncertainty. The distances are taken
from \citet{1997ChristianSwank} unless indicated otherwise. \citeauthor{1997ChristianSwank}
take the uncertainty in the distance to be 30\%. \citet{2004Jonker}
indicate that the uncertainty in the distances they determined is
{}``probably very large''. Therefore we use the 30\% from \citeauthor{1997ChristianSwank}.
The uncertainty in the luminosity is dominated by the uncertainty
in the distance. Since $L\propto d^{2}$, the uncertainty in $L$
is 60\%. As the ignition column depth is proportional to the luminosity
(see equation \ref{eq:y11}), it has the same uncertainty of 60\%.
In the case of GX~17+2 this has to be added to the indicated uncertainty
in $y$ from the spread in the observed recurrence time.}}

\begin{center}\begin{tabular}{llllll}
\hline 
Source&
$F$ {[}{\small $\mathrm{10^{-8}erg\, s^{-1}cm^{-2}}$}{]}&
variability&
$d$ {[}kpc{]}&
$L/L_{\mathrm{Edd}}$&
$y$ {[}$10^{11}\mathrm{g\, cm^{-2}}${]}\tabularnewline
\hline
Sco X-1&
&
13\%&
$2.8\pm0.3^{\mathrm{a}}$&
$0.9\pm0.3^{\mathrm{b}}$&
>1.8\tabularnewline
GX 340+0&
$2.67^{\mathrm{c}}$&
15\%&
<$11$&
<1.9&
>9.4\tabularnewline
GX 349+2&
$2.45^{\mathrm{d}}$&
16\%&
$5.0$&
0.4&
>1.8\tabularnewline
GX 9+9&
$0.8\pm0.012^{\mathrm{e}}$&
14\%&
$5.0$&
0.1&
>0.6\tabularnewline
GX 354-0&
$0.037^{\mathrm{f}}$&
37\%&
$4.5^{\mathrm{g}}$&
0.05&
>0.2\tabularnewline
GX 5-1&
$2.56\pm0.4^{\mathrm{h}}$&
8\%&
<$9.0$&
<1.2&
>4.4\tabularnewline
GX 9+1&
$2.1^{\mathrm{i}}$&
13\%&
$5^{\mathrm{i}}$&
0.3&
>1.0\tabularnewline
GX 13+1&
$0.803\pm0.004^{\mathrm{j}}$&
7\%&
$7.0$&
0.2&
>0.7\tabularnewline
GX 17+2&
$2.26^{\mathrm{k}}$&
12\%&
$7.5$&
0.8&
$1.6\pm0.8$\tabularnewline
Cyg X-2&
$1.3^{\mathrm{l}}$&
23\%&
$8.0$&
0.5&
>2.0\tabularnewline
\hline
\end{tabular}\end{center}

$^{\textrm{a}}$ {\small Distance from parallax measurements (\citealt{1999Bradshaw})}{\small \par}

$^{\mathrm{b}}$ {\small From \citet{2003Bradshaw} (2.0--18.2 keV)}{\small \par}

{\small $^{\textrm{c}}$ From \citet{2004Lavagetto} (0.1--200 keV)}{\small \par}

{\small $^{\textrm{d}}$ From \citet{2004Iaria} (0.1--200 keV)}{\small \par}

{\small $^{\textrm{e}}$ From observations with BeppoSAX NFI (0.1--200
keV). See Table \ref{cap:Flux-determination-with}.}{\small \par}

{\small $^{\textrm{f}}$ From \citet{2003Galloway} (2--60 keV)}{\small \par}

{\small $^{\textrm{g}}$ Distance from \citet{2004Jonker} for hydrogen
accreting neutron star.}{\small \par}

{\small $^{\textrm{h}}$ From \citet{2003A&A...411L.363P} (1.5--40
keV). Note that the flux in the 12--20 keV range is obtained through
linear interpolation.}{\small \par}

{\small $^{\textrm{i}}$ From \citet{2005Iaria2005} (0.12--18 keV)}{\small \par}

{\small $^{\textrm{j}}$ From \citet{2003Corbet} (1.5--12 keV) and
\citet{Revnivtsev:2004sg} (18--60 keV). Flux in 12--18 keV range
obtained through linear interpolation. Total energy range: 1.5--60
keV.}{\small \par}

{\small $^{\textrm{k}}$ From \citet{2004intZand} (0.1--200 keV)}{\small \par}

{\small $^{\textrm{l}}$ From \citet{2002DiSalvo} (0.1--200 keV)}{\small \par}

\end{table*}

A measurement of the mass accretion rate is most directly done through
the X-ray luminosity because the power for the radiation is provided
by the liberation of gravitational energy of the accreted matter.
However, this is not unambiguous because it is somewhat uncertain
what fraction of the liberated energy goes into radiation. It is generally
thought that some fraction goes into the kinetic energy of jets. Nevertheless,
in the case of neutron star accretors it is not unreasonable to assume
that most of the energy goes into radiation (see e.g. \citealt{2002Fender}).

Another complication involves the determination of the luminosity:
distances to LMXBs --- needed to translate fluxes to luminosities
--- are generally poorly known (except for systems in globular clusters).
This applies particularly to many of the systems that we study in
this paper: persistent LMXBs outside globular clusters without type-I
X-ray burst behavior. Systems that exhibit bursts with Eddington-limited
fluxes, as diagnosed through photospheric radius expansion, provide
a means to find distances with $\approx30\%$ accuracy (cf., \citealt{2003Kuulkers}),
and transients that turn quiescent may reveal the optical light of
the companion star without contamination by optical light from the
accretion disk. Neither of these two conditions apply to most of our
systems.

A third complication is that for many systems we do not have a broad-band
measurement of the X-ray spectrum that is averaged on timescales longer
than that of the typical variability to enable a good bolometric correction.
Broadband BeppoSAX X-ray spectra have been taken in many cases but
almost never for long enough. This situation may change soon with
the results of the deep exposures of INTEGRAL on the Galactic bulge.

We attempted to obtain the best broadband flux measurements for the
ten systems here studied, see Table \ref{cap:Results-for-non-bursting}.
For nine sources this pertains to values found in the literature from
RXTE, BeppoSAX and INTEGRAL observations, and for one system, GX~9+9,
we perform the analysis ourselves from BeppoSAX data. The data were
taken on April 8, 2000, with exposure times of 23, 50 and 22 ks for
LECS, MECS and PDS, respectively. Background spectra are taken from
blank field observations for LECS and MECS, at the same detector positions,
and from off-source pointings with the PDS. We rebin the spectra to
limit the oversampling of the spectral resolution to 40\% of the full-width
at half maximum, and to obtain at least 15 photons in each spectral
bin. During the spectral analysis, a systematic error of 1\% is added
in quadrature to the statistical error per spectral bin and the normalization
between the three instruments is allowed to vary. We find the generic
LMXB spectral model to provide good fits to the data. This model (see
e.g. \citealt{2001Sidoli}) consists of a multi-temperature disk black
body (\citealt{1984Mitsuda,1986Makishima}) in combination with a
comptonized spectrum (\citealt{1994Titarchuk,1995Hua,1995Titarchuk}),
both absorbed by cold interstellar matter following the model by \citet{1983Morrison}.
The results of the fit are provided in Table \ref{cap:Flux-determination-with}.%
\begin{table}

\caption{\label{cap:Flux-determination-with}Flux of GX~9+9 from BeppoSAX
NFI observations. {\small As model we use an absorbed disk black body
in combination with a comptonized spectrum. We provide the parameters
from the best fit of the model to the observed spectrum and the unabsorbed
0.1--200~keV flux.}}

\begin{center}\begin{tabular}{ll}
\hline 
$N_{\mathrm{H}}$&
$\left(0.231\pm0.003\right)10^{22}\,\mathrm{cm^{-2}}$\tabularnewline
&
\tabularnewline
$kT_{\mathrm{BB}}$&
$0.853\pm0.002$ keV\tabularnewline
$\mathrm{Flux}_{\mathrm{BB}}$(0.1--200 keV)&
$\left(0.33\pm0.014\right)10^{-8}\,\mathrm{erg\, s^{-1}\, cm^{-2}}$\tabularnewline
&
\tabularnewline
$kT_{0}$&
$1.055\pm0.003$ keV\tabularnewline
$kT_{e}$&
$2.458\pm0.004$ keV\tabularnewline
$\tau$&
$14.10\pm0.06$\tabularnewline
$\mathrm{Flux}_{\mathrm{comptt}}$(0.1--200 keV)&
$\left(0.50\pm0.04\right)10^{-8}\,\mathrm{erg\, s^{-1}\, cm^{-2}}$\tabularnewline
&
\tabularnewline
$\chi^{2}/\mathrm{dof}$&
$166/147$\tabularnewline
Flux (0.1--200 keV)&
$\left(0.83\pm0.013\right)10^{-8}\,\mathrm{erg\, s^{-1}\, cm^{-2}}$\tabularnewline
\hline
\end{tabular}\end{center}
\end{table}

The flux measurements are based on observations which were typically
performed during 1.3 days. From the one-day averaged ASM/RXTE lightcurves
we determine the fractional rms variability for each source, see Table
\ref{cap:Results-for-non-bursting}. We applied the best distance
estimates from the literature and determined the ratio of the resulting
luminosity and the Eddington luminosity of a canonical neutron star
(i.e. $L_{\mathrm{Edd}}\equiv2\cdot10^{38}\mathrm{erg}\,\mathrm{s}^{-1}$
for a neutron star with a mass of $1.4\, M_{\odot}$, a 10~km radius
and a hydrogen-rich photosphere). Due to the three sources of uncertainty
eluded to above, the resulting ratios are never very accurate, and
we are only able to make a crude distinction between the sources close
to one tenth and 100\% of the Eddington limit. This inference is in
line with the fact that the five sources close to the Eddington limit
are the only ones that trace out a Z-shaped curve in a hardness-intensity
diagram (e.g., \citealt{1989Hasinger}). They are Sco~X-1, GX~340+0,
GX~5-1, GX~17+2 and Cyg~X-2. The others are so-called Atoll sources.

\section{Discussion}

We analyzed a large volume of data from the Wide Field Cameras on
BeppoSAX of ten luminous LMXBs to search for superbursts. Except for
those already reported from GX~17+2 by \citet{2004intZand}, none
were found and we set lower limits on the recurrence time of superbursts
that range from 30 to 76 days.

One of the conditions for a superburst to occur is a sufficiently
large column depth of accumulated matter. The ignition column depth
$y$ for a $1.4\, M_{\odot}$ neutron star with a 10 km radius, assuming
a gravitational redshift $z=0.31$, is given by (\citealt{2004intZand})
\begin{equation}
y=2.0\cdot10^{11}\mathrm{g\, cm^{-2}}\left(\frac{t_{\mathrm{recur}}}{30\,\mathrm{days}}\right)\left(\frac{\dot{m}}{\dot{m}_{\mathrm{Edd}}}\right),\label{eq:y11}\end{equation}
with $\dot{m}_{\mathrm{(Edd)}}$ the (Eddington limited) mass accretion
rate. At a given accretion rate, the lower limit to the recurrence
time gives a lower limit to the ignition column depth. Using Equation
(\ref{eq:y11}) we calculate $y$ (see Table \ref{cap:Results-for-non-bursting}).

\begin{figure}
\begin{center}\includegraphics[%
  clip,
  width=1.0\columnwidth]{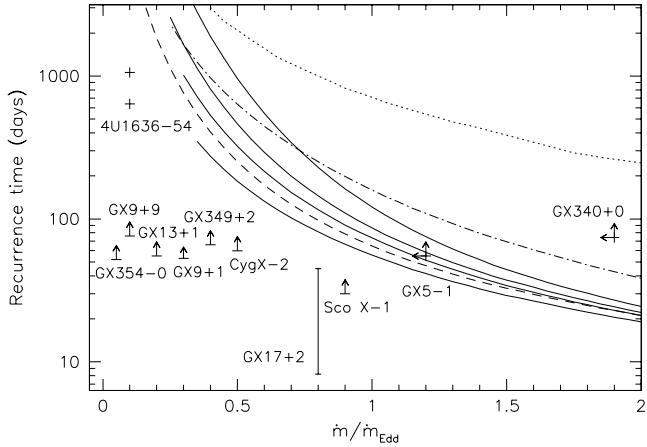}\end{center}

\caption{\label{cap:Observed-recurrence-time}{\small Superburst recurrence
time as a function of the mass accretion rate in units of the Eddington
limited mass accretion rate. We show models from \citet{2005Cumming1}
and observed recurrence times including our lower limits. The solid
curves show results for a model with a disordered crust, a composition
of $^{\textrm{56}}$Fe and a 20\% carbon abundance, without Cooper
pairing for four different core neutrino emissivities. The dashed
curve is for a model with heavier composition ($A=104$), the dot-dashed
curve has a higher crust conductivity and the dotted curve includes
Cooper pair neutrino emission. The curves end at lower $\dot{m}$
where stable carbon burning begins.}}

{\small Note that the uncertainty in the mass accretion rate is at
least 60\% for most sources (see Table \ref{cap:Results-for-non-bursting}).}
\end{figure}

\citet{2005Cumming1} investigate several carbon ignition models for
superbursts. From these the recurrence time as a function of the mass
accretion rate is determined. In Figure \ref{cap:Observed-recurrence-time}
we compare our lower limits to $t_{\mathrm{recur}}$ as well as the
observed recurrence times of 4U 1636-536 and GX~17+2 to these models.
Looking at what theory predicts, the constraints that we can put on
the superburst recurrence time are only meaningful for sources that
accrete close to the Eddington limit. For the sources that accrete
close to 10\% of the Eddington limit the theory predicts for all models
a higher $t_{\mathrm{recur}}$ than our lower limit. Sco~X-1, GX~340+0,
GX~5-1 and Cyg~X-2 accrete matter at a rate comparable to GX~17+2,
namely close to the Eddington limit.

\citet{2005Cumming1} discuss the dependence of the ignition of carbon
on the heat flux into the superbursting layer from the deeper lying
crust: $Q_{b}$. At a given accretion rate, the recurrence time is
lower for higher $Q_{b}$. Therefore a lower limit to $t_{\mathrm{recur}}$,
or alternatively to $y$, gives an upper limit to $Q_{b}$ for a given
model (see Figure 20 in \citet{2005Cumming1}). Comparing the lower
limit to the ignition column depth of the five brightest sources to
the results of \citet{2005Cumming1} we find $Q_{b}\lesssim0.1$~MeV
per nucleon for iron composition. However, the large uncertainty in
the accretion rate due to the large uncertainty in the distance to
these sources prevents us from placing stringent constraints on the
models of \citet{2005Cumming1}.

We found the lower limit to the superburst recurrence time for GX~340+0,
GX~5-1 and Cyg~X-2 to be about two times the recurrence time of
GX~17+2. One of our goals was to learn more about the dependence
of superbursts on normal bursting behavior. However, we cannot explain
the difference in $t_{\mathrm{recur}}$ by the difference in normal
type I bursting behavior as both GX~17+2 and Cyg~X-2 exhibit bursts,
while GX~340+0 and GX~5-1 do not.

From Figure \ref{cap:Observed-recurrence-time} we see that, depending
on the model, a 35\% difference in the mass accretion rate can lead
to a two times longer recurrence time. The uncertainty in the mass
accretion rate is 60\%. Therefore the difference in (lower limit on)
$t_{\mathrm{recur}}$ could be due to GX~17+2 accreting faster than
GX~340+0, GX~5-1 and Cyg~X-2.

Another possible explanation for the difference in $t_{\mathrm{recur}}$
is that GX~17+2 may differ from the other sources in some aspect
which would lower the theoretical curves for this source. The core
temperature of GX~17+2 might be higher, which would lead to a lower
ignition column depth and therefore a lower recurrence time (see e.g.
Figure 11 in \citealt{2005Cumming1}). Alternatively there may be
a difference in the hydrogen and helium abundance of the accreted
matter. A larger initial helium abundance will lead directly to a
larger carbon yield from rp-process burning (\citealt{2003Schatz}).
\citet{2001CummingBildsten} find that the amount of carbon required
to achieve unstable ignition at accretion rates close to the Eddington
limit is approximately 3--5\%. The presence of observable superbursts
in GX~17+2 might indicate that the accreted material in this source
is hydrogen deficient, allowing sufficient carbon to be made, whereas
too little carbon is made in other sources to result in a superburst.
Another effect, which goes in the opposite direction, is that a lower
initial hydrogen abundance will lead to a less extensive rp-process,
giving less massive heavy elements, which increases the carbon ignition
depth because of the lower opacity of the accumulating fuel layer
(\citealt{2001CummingBildsten}). However, \citet{2004Brown} shows
that this effect is smaller than initially thought. At present only
one superburster, the ultracompact system \object{4U~1820-30}, is assumed
to accrete hydrogen-poor matter. In this case, the extremely short
orbital period of only 11.4 minutes (\citealt{1987Stella}) implies
that the hydrogen mass fraction is $X\lesssim10$\%, consistent with
the very luminous and energetic superburst observed from this source
(\citealt{2003Cumming}). Another case of a superburster accreting
hydrogen-deficient matter is possibly the ultracompact system \object{4U~0614+091}
(see e.g. \citealt{2004Nelemans}) of which most recently a superburst
was observed (\citealt{2005ATel..483....1K}).

\section{Conclusion}

We searched for superbursts from ten bright LMXBs. None were found
but the four known superbursts from GX~17+2. For the other sources
we obtain lower limits to the recurrence time and the ignition column
density. Comparing this lower limit for the five most luminous sources,
Sco~X-1, GX~340+0, GX~5-1, GX~17+2 and Cyg~X-2, to the models
by \citet{2005Cumming1} $Q_{b}\lesssim0.1$~MeV per nucleon is required.
Due to the large uncertainty in the distance to these sources, the
uncertainty in the mass accretion rate is large as well. This prevents
us from placing firm constraints on the predictions by \citet{2005Cumming1}.
As our search method is indiscriminate of the burst ignition scenario,
the recurrence time limits may also be applied to other bursts of
similar duration and brightness.

The four luminous non-superbursters have a mass accretion rate comparable
to the luminous superburster GX~17+2. For three of these sources
we find a lower limit to the recurrence time that is twice as long
as the observed average $t_{\mathrm{recur}}$ of GX~17+2. Most likely
this difference is due to GX~17+2 accreting faster than the other
sources and therefore GX~17+2 has a lower $t_{\mathrm{recur}}$.
However, we are unable to determine whether this is the case because
of the large uncertainty in the mass accretion rate.

We found no superbursts in the six non-bursting sources we considered,
which confirms the current observational view of superbursters, i.e.
all known superbursters exhibit normal type I X-ray bursts as well.
This supports the suggestion that both stable and unstable burning
of hydrogen and helium are necessary to produce the fuel for superbursts.

To better constrain models of neutron star interior physics, more
long term monitoring observations of (candidate) superbursters are
necessary. As superbursts have a long recurrence time, a long exposure
time is required for each observed source. The planned mission MIRAX
(\citealt{2004Braga}) will be the most suited for such a task. MIRAX
will have on board the spare flight unit of the BeppoSAX WFCs as well
as two hard (10--200 keV) X-ray cameras. The aim of MIRAX is to perform
continuous broadband observations of the Galactic center during nine
months per year. This will provide a high probability to observe any
superbursts that have a recurrence time less than approximately these
nine months, which includes the predicted recurrence time of many
models shown in Figure \ref{cap:Observed-recurrence-time} for sources
which accrete close to the Eddington limit. The latest observations
of superbursts (\citealt{2005ATel..482....1R,2005ATel..483....1K})
were done with RXTE/ASM, which observes 80\% of the sky each 90 minutes.
Due to a time resolution of 90~s the short (1--10~s) burst rise
cannot be observed, which makes it more difficult to distinguish between
a superburst and a flare. Observations from MIRAX will have a time
resolution better than 0.5~ms which is more than adequate to observe
the rise of superbursts.

\begin{acknowledgements}
We thank M.~Méndez for his useful comments on an earlier version
of this paper. We made use of quick-look results provided by the ASM/RXTE
team. SRON is financially supported by the Netherlands Organization
for Scientific Research (NWO). AC acknowledges support from McGill
University startup funds, an NSERC Discovery Grant, Le Fonds Québécois
de la Recherche sur la Nature et les Technologies, and the Canadian
Institute for Advanced Research.
\end{acknowledgements}
{\small \bibliographystyle{aa}
\bibliography{4884}
}
\end{document}